\documentclass[preprint,showpacs,preprintnumbers,amsmath,amssymb]{revtex4}
\usepackage{graphicx, epsfig, dcolumn}
\begin{document}


\title{Factors influencing the Martensitic interactions in Ni$_{50}$Mn$_{35}$Sn$_{15}$: An EXAFS study}
\author{P. A. Bhobe} 
\author{K. R. Priolkar}\email[corresponding author: ]{krp@unigoa.ac.in} \author{P. R. Sarode}
\affiliation{Department of Physics, Goa University, Goa, 403 206 India.}

\date{\today}

\begin{abstract}
Extended x-ray absorption fine-structure (EXAFS) measurements have been carried out on Ni$_{50}$Mn$_{35}$Sn$_{15}$ in the austenitic and martensitic phase. The temperatures associated with structural and magnetic phase transformations are obtained from magnetization measurements. The system orders ferromagnetically below 319 K while the structural phase transition occurs at T$_M \sim$ 200 K. EXAFS measurements have been carried out at Mn and Ni K-edges and changes associated with respect to the local structure of these absorbing atoms are compared. Significant changes in the nearest neighbour interactions are seen as the system transits to the low temperature martensitic phase. EXAFS provides evidence for local structural distortion even in the cubic phase that is not seen from the x-ray diffraction studies. The results obtained are understood in terms of changing hybridization between the constituent atoms that influence the electronic structure at the Fermi level, associated with the austenitic-to-martensitic transition. 

\end{abstract}
\maketitle

\section{\label{sec:level1}Introduction}
Martensitic transformations and its pairing with ferromagnetism has been a central subject for investigation in the recent years. Especially, some intermetallics show simultaneous occurrence of martensitic and magnetic transitions, suggesting the possibility of controlling the structural transformation by magnetic field and could be exploited for practical applications. Such multifunctional materials are classified under rapidly growing technological field of Ferromagnetic Shape Memory alloys (FSMA). Among the variety of FSMA, Ni-Mn-Ga alloys are a recently synthesized class of alloys that have been studied extensively and hence serve as a reference in the development of new systems \cite{vasil1, sod, ent}. The stoichiometric Ni$_2$MnGa undergoes martensitic transition around 220 K from a L2$_1$ cubic phase to a low symmetry modulated structure, while the ferromagnetic transition takes place at 370 K \cite{web}. An interesting aspect of Ni-Mn-Ga alloys is the isothermal giant entropy change obtained when structural and magnetic transition temperatures nearly coincide, leading to a development of new materials exhibiting magnetocaloric effect \cite{pareti}. The latest candidates in the field of FSMA has been alloys with composition Ni$_{50}$Mn$_{50-x}$Z$_{x}$ with Z = In, Sn, Sb \cite{suto}. Of particular interest is Ni$_{50}$Mn$_{50-x}$Sn$_{x}$ with 0.13$\le x \le$0.15 for which an inverse magnetocaloric effect is observed that is nearly three times larger in comparison to other alloys \cite{acet-nature}. This system orders ferromagnetically at a Curie temperature T$_C$ $\sim$ 319 K while martensitic transition occurs at temperature, T$_M \sim $ 200 K \cite{suto}. Neutron diffraction experiments on Ni$_{50}$Mn$_{36}$Sn$_{14}$ show that the cubic L2$_1$ structure in the austenitic phase transforms to orthorhombic 4$O$ structure with {\it Pmma} space group in the martensitic phase \cite{brown18}. Studies show that despite structural similarity with Ni$_2$MnZ alloys, strong differences in the underlying martensitic and magnetic interactions are seen in Ni$_{50}$Mn$_{50-x}$Sn$_{x}$ \cite{acet, koya88}. For example, while the martensitic transition is absent in the stoichiometric Ni$_2$MnSn, these Mn rich alloys exhibit martensitic transition in the wide range of temperature \cite{suto}. Further, the magnetic moment of Ni$_{50}$Mn$_{36}$Sn$_{14}$ in the martensitic phase is smaller by about 50\% than that in cubic phase \cite{koya88}. Even in the L2$_1$ phase, the Mn moments are significantly smaller than those reported for the stoichiometric Ni$_2$MnZ alloys \cite{zaybk,brown18}. It is conjectured that apart from the ferromagnetic order, some antiparallel alignment of the excess Mn moments could exist in Ni$_{50}$Mn$_{50-x}$Sn$_{x}$ \cite{acet, brown18}. Thus an understanding of the local structure in the martensitic and austenitic phases is vital in establishing a complete picture of the transformation mechanics. Therefore, we have carried out {\it extended x-ray absorption fine structure} (EXAFS) measurements at Mn and Ni K-edges in the austenitic and martensitic phases of Ni$_{50}$Mn$_{35}$Sn$_{15}$.

\section{\label{sec:level1}Experimental Details}
\begin{figure}[h]
\epsfig{file=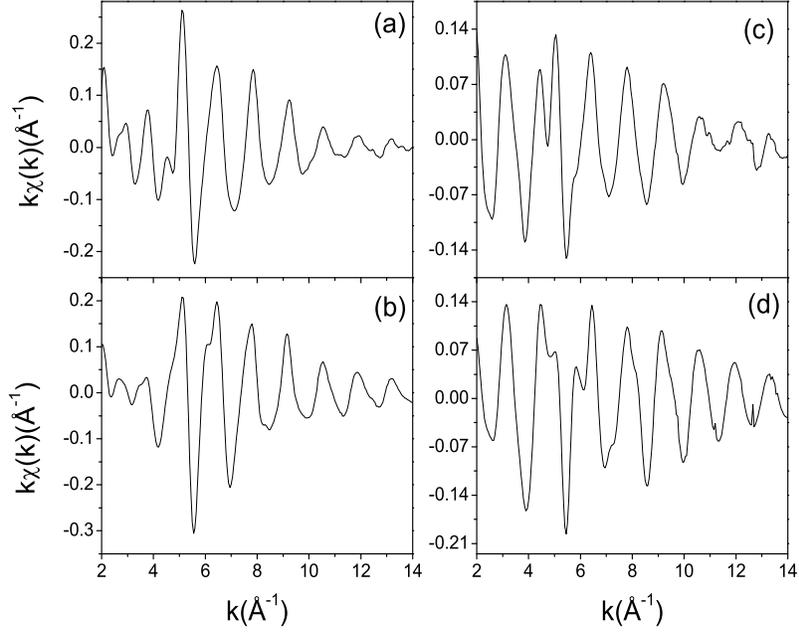, width=12cm, height=10cm}
\caption{\label{raw-xafs}The $k$-weighted $\chi (k)$ spectra of Ni$_{50}$Mn$_{35}$Sn$_{15}$ recorded at (a) room temperature Mn K edge (b) low temperature Mn K edge (c) room temperature Ni K edge and (d) low temperature Ni K edge. These data were Fourier transformed in the range (2-12) (\AA$^{-1}$) for analysis.}
\end{figure}

A homogeneous bead of Ni$_{50}$Mn$_{35}$Sn$_{15}$ was prepared by repeated melting of the appropriate quantities of the constituent elements of 4 N purity under argon atmosphere in an arc furnace. The sample bead so obtained was sealed in an evacuated quartz ampoule and annealed at 800 K for 48 h followed by quenching in cold water. Energy dispersive x-ray analysis that gave the elemental content as: Ni = 50.4, Mn = 34.6  and Sn = 15.0 confirmed the composition of the sample to be close to nominal. The sample bead was cut and thoroughly ground to a very fine powder for x-ray diffraction (XRD) and  EXAFS measurements while a small piece of the same bead was used for magnetization study.  The room temperature crystal structure was determined by XRD recorded on Rigaku D-MAX IIC diffractometer with Cu K$\alpha$ radiation. The magnetization measurements were carried out on a Vibrating Sample Magnetometer in the low field value (50 Oe) and in the temperature range 50 to 350 K. EXAFS at Ni and Mn K-edge were recorded at room temperature and liquid nitrogen temperature in the transmission mode on the EXAFS-1 beamline at ELETTRA Synchrotron Source using Si(111) as monochromator. For this, the powder of Ni$_{50}$Mn$_{35}$Sn$_{15}$ was coated on scotch-tape strips. These sample coated strips were adjusted in number such that the absorption edge jump gives $\Delta\mu x \le 1$. The incident and transmitted photon energies were simultaneously recorded using gas-ionization chambers as detectors. Measurements were carried out from 300 eV below the edge energy to 1000 eV above it with a 5 eV step in the pre-edge region and 2.5 eV step in the EXAFS region. At each edge, three scans were collected. Data analysis was carried out using IFEFFIT \cite{new}  in ATHENA and ARTEMIS programs \cite{brav}. Here theoretical fitting standards were computed with ATOMS and FEFF6 programs \cite{rav, zab}. The $k$-weighted $\chi(k)$ spectra at Mn and Ni K-edges at room temperature (RT) and liquid nitrogen temperature (LT) are shown in Fig.\ref{raw-xafs}. It can be seen that there are distinct differences in the RT and LT spectra, especially in the k range of 4 to 7\AA$^{-1}$ which can be ascribed to martensitic transformation of the sample.  The data in the $k$ range of (2 - 12)\AA$^{-1}$ was used for analysis. For XANES, the data was collected from 300 eV below the edge energy to 100 eV above the edge in steps of 5 eV from -300 eV to -50 eV; 0.2 eV from -50 eV to +50 eV and 1 eV for the rest of spectra. Normalization was done by first subtracting the Victoreen instrument background obtained from fitting the pre-edge
region (-200 eV to -50 eV) from the all of the raw spectra and then dividing them by respective average absorption coefficients obtained for the spectral region +50 eV to +90 eV.

\section{\label{sec:level1}Results}
\begin{figure}[h]
\epsfig{file=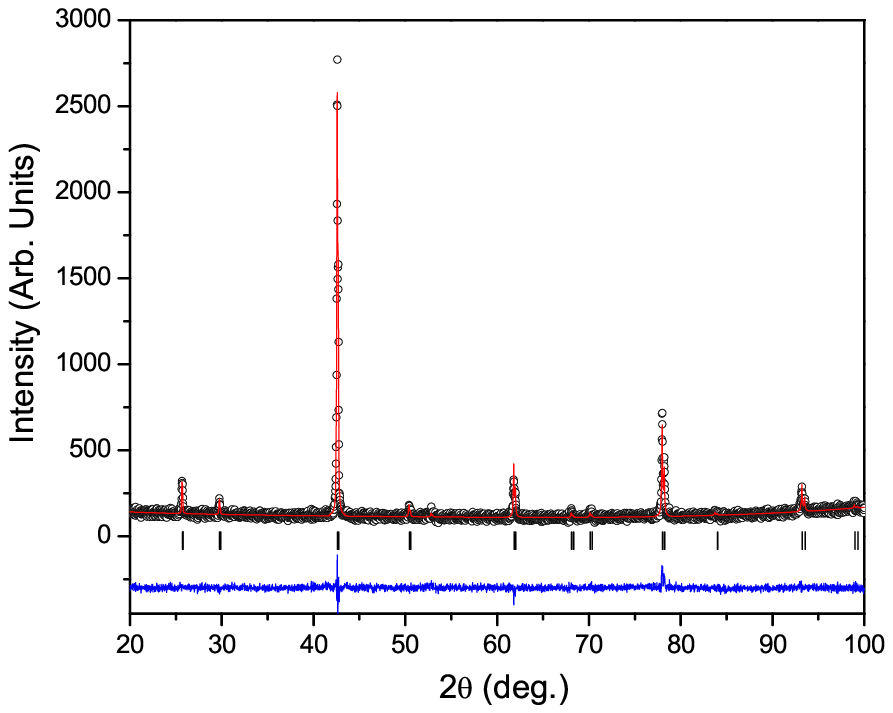, width=10cm, height=8cm}
\caption{\label{sn-xrd}(Color online) The x-ray powder diffraction pattern of Ni$_{50}$Mn$_{35}$Sn$_{15}$ at room temperature. The open circles show the observed counts and the continuous line passing through these counts is the calculated profile. The difference between the observed and calculated patterns is shown as a continuous line at the bottom of the two profiles. The calculated positions of the reflections are shown as vertical bars.}
\end{figure}

Rietveld refinement of the room temperature XRD data confirms that the sample is highly ordered in the Heusler L2$_1$ structure with {\it Fm3m} space group and lattice parameters of 5.9941$\pm$0.0003 \AA. The observed and calculated diffraction patterns together with the difference pattern are shown in Fig. \ref{sn-xrd}. In the stoichiometric Heusler composition X$_2$YZ, Ni$_{50}$Mn$_{35}$Sn$_{15}$ can be described as X = Ni occupying the ($\frac{1}{4} \frac{1}{4} \frac{1}{4}$) site, Y = Mn occupying the ($0 0 0$) site and Z = (60\% Sn + 40\% Mn) occuping the ($\frac{1}{2} \frac{1}{2} \frac{1}{2}$) site. To distinguish between the Mn occupying the (0, 0, 0) position and the excess Mn that occupies the Z sites, it will henceforth be refered to as Y-Mn and Z-Mn respectively. 

\begin{figure}[h]
\epsfig{file=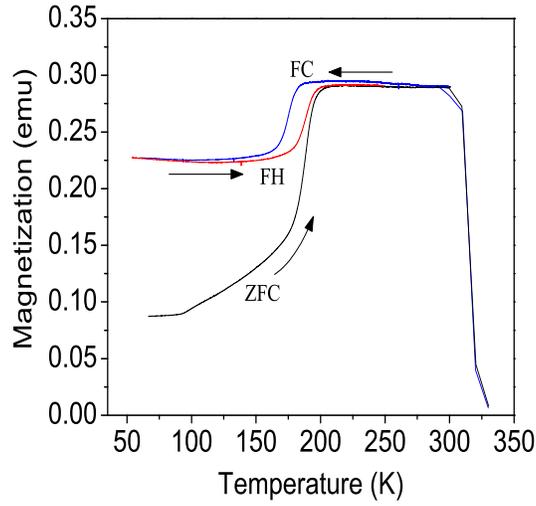, width=8cm, height=8cm}
\caption{\label{sn-mag}(Color online) Temperature dependence of magnetization of Ni$_{50}$Mn$_{35}$Sn$_{15}$ measured in an applied field of 50 Oe between 50 to 350 K. Measurements were carried out while heating (ZFC), cooling (FC) and re-heating (FH) the sample as indicated by the arrows. The structural and magnetic transition temperatures were determined from the differentials of magnetization with respect to temperature for FC and FH curves.}
\end{figure}

Magnetization was determined in a field of 50 Oe over the temperature range 50-350 K. Initially, the sample was cooled from room temperature to 50 K in the absence of magnetic field. An external field of 50 Oe was applied and magnetization was recorded with increasing temperature upto 350 K (referred as ZFC). The measurements were made as the sample was subsequently cooled to 50 K without disturbing the magnetic field (FC) and finally the sample was field-heated (FH) upto 350 K. The ferromagnetic ordering sets in at Curie temperature T$_C$ = 319 K. As the temperature is lowered, magnetization remains essentially constant until 200 K where an abrupt fall in $M$ is observed. This behaviour corroborates with that reported for the martensitic phase transition by Ref. 6, 8, 9 and 12 in Ni-Mn-Sn alloys. The ZFC, FC and FH magnetization curves are shown in Fig. \ref{sn-mag}. The martensitic start (M$_s$ = 182.7 K), martensitic finish (M$_f$ = 164.3 K), reverse martensitic or austenitic start (A$_{s}$ = 176.2 K) and austenitic finish (A$_{f}$ = 195.6 K) temperatures are identified and the thermal hysteresis is found to be $\sim$12 K. The hysteresis in the FC and FH curves is attributed to the structural transition while the splitting in the ZFC and FC curves just below the T$_C$ has been associated with competing magnetic interactions speculated in this system \cite{acet}. The presence of Z-Mn in the system give rise to additional Mn-Mn interactions with antiparallel alignment of moments that is responsible for the lower magnetic moments in the ausenitic as well as martensitic phases. 

\begin{figure}[h]
\epsfig{file=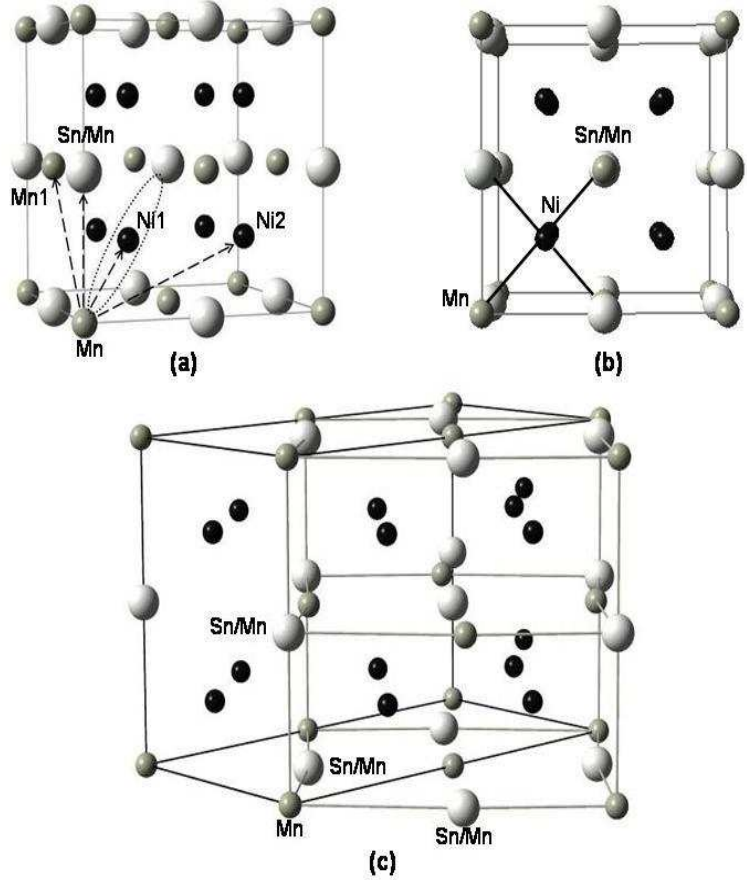, width=12cm, height=12cm}
  \caption{(Color online) The crystal structure of Ni$_{50}$Mn$_{35}$Sn$_{15}$ based on which the EXAFS have been modeled. (a) The L2$_1$ cubic phase shows the absorbing atom and the relevant backscatters. The arrows indicate different SS paths while the dotted loop shows the MS path. (b) The plane of atoms in the cubic crystal represents the fact that Y-Mn and Sn/Z-Mn are equidistant from Ni. (c) The {\it Pmma} crystal structure in the martensitic phase wherein the L2$_1$ type atomic arrangement is visualized.}
  \label{cube}
\end{figure}

Room temperature EXAFS recorded at Mn K-edge are presented in Fig. \ref{raw-xafs}(a). With the knowledge of crystal structure from XRD, a cubic L2$_1$ model with $a$ = 5.994\AA~ was adapted for the analysis of this data. Here, the most relavent contribution to EXAFS in the $R$-range 1.0 - 5.0 \AA, ~comes from near neighbour interactions that comprises of four single scattering (SS) paths and one linear multiple scattering (MS) path along the body diagonal of the initial cubic cell as shown schematically in Fig. \ref{cube}(a). As per the L2$_1$ model, the second SS path is entirely due to scattering from Z atoms. Due to the presence of excess Mn in Ni$_{50}$Mn$_{35}$Sn$_{15}$, the Z-site is occupied by two different type of atoms, Sn and Mn. Inorder to incorporate the contribution from both these atoms occupying the Z-site, two fitting standards were calculated in the FEFF input file. The first fitting standard was calulated with Sn atoms alone occupying the Z site, while in the second, only Mn atoms were considered to be entirely occupying the Z site. Both these fitting standards were then introduced in the model with respective path coordination numbers changed as per the actual composition ratio (Sn:Mn = 60:40). The structure being cubic at room temperature, the correction to the path lengths was refined with a constraint, $\delta R$ = $\delta r_1$ $\times$ $[R_{eff}/\ R_{nn1}]$ where $R_{nn1}$ is the nearest neighbour distance, kept fixed to 2.5955\AA~ (calculated from the lattice constant), $R_{eff}$ is the calculated bond length obtained from FEFF and $\delta r_1$ is the change in first neighbour distance. The thermal mean-square variation in the bond lengths, $\sigma^2$ were varied independently for each path considered in the fit. Such an approach of data analysis has been successfully used for the EXAFS of Ni$_{2+x}$Mn$_{1-x}$Ga alloys \cite{pab}. The present fit gives the first neighbour, Mn-Ni distance equal to 2.549\AA~ which is much shorter than 2.5955\AA, the value calculated from the lattice constant. This points towards some disorder at the local level in the cubic structure. However, no structural disorder is seen either in the XRD profile (refer Fig. \ref{sn-xrd}) or in the neutron diffraction study by Brown {\it et.al} \cite{brown18} on Ni$_{50}$Mn$_{36}$Sn$_{14}$ system. 

\begin{figure}[h]
\epsfig{file=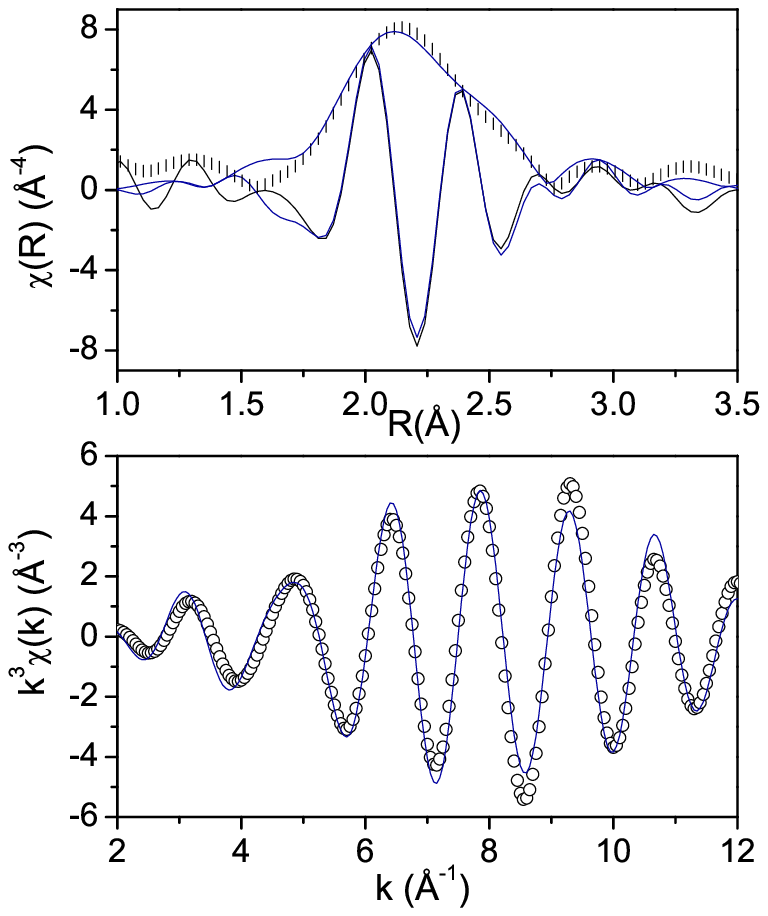, width=8cm, height=10cm}
\caption{\label{sn-nirt}(Color online) Magnitude and real component of FT of EXAFS spectra in R space (top panel) and real component of FT in the back transformed k space (bottom panel) for Ni K-edge in Ni$_{50}$Mn$_{35}$Sn$_{15}$ obtained at room temperature. The fitting to the data are shown as coloured lines.}
\end{figure}

In the case of Ni K-edge EXAFS the fitting was carried out in the $R$-range 1.0 - 3.0 \AA~. The peak in this range of the spectra comprises of Ni-Mn, Ni-Sn and Ni-Ni single scattering paths. The analysis similar to that of Mn EXAFS, with constrainted refinement of $\delta R$ was carried out. However, this did not result in a good fit. To overcome this situation, $\delta r$s' for each of the three scattering paths were varied independently. As can be seen in Fig. \ref{sn-nirt}, a good fit was obtained for values listed in Table \ref{tab:sn-aus}. It is interesting to note that Ni-Mn bond length is distinctly shorter than Ni-Sn bond length. Further, the Ni-Sn and Ni-Ni bond lengths match closely with the values calculated from the room temperature XRD data. 

\begin{figure}[h]
\epsfig{file=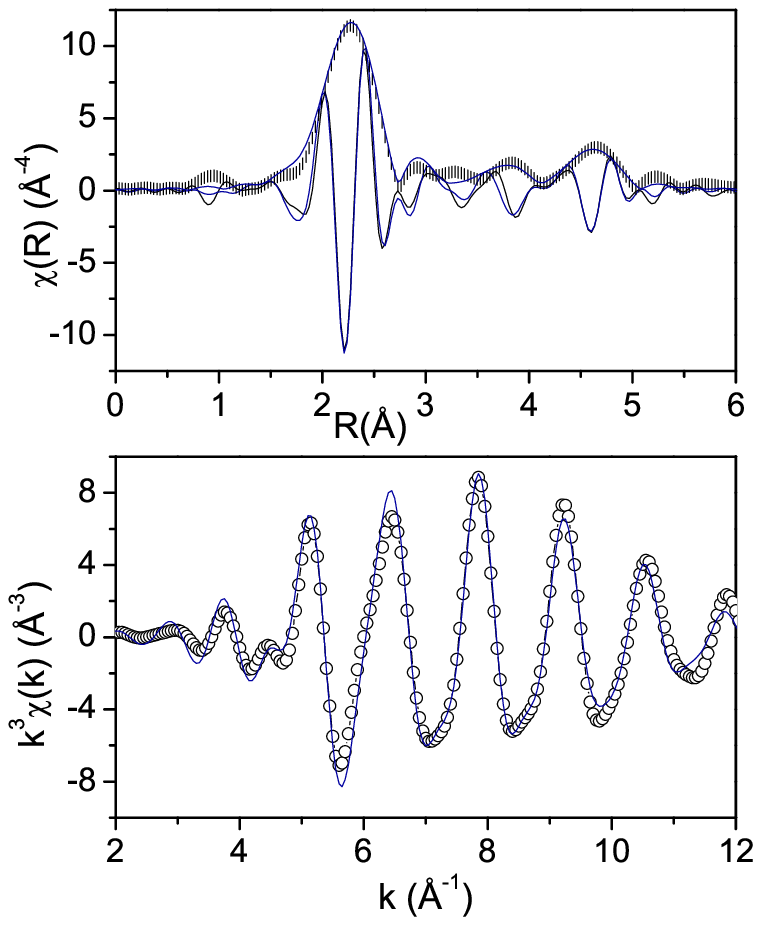, width=8cm, height=10cm}
\caption{\label{sn-mnrt}(Color online) Magnitude and real component of FT of EXAFS spectra in R space and real component of FT in the back transformed k space for Mn K-edge in Ni$_{50}$Mn$_{35}$Sn$_{15}$ obtained at room temperature. The fitting to the data are shown as coloured lines.}
\end{figure}

In the cubic L2$_{1}$ structure, the X-site(Ni) is surrounded by four Y-site(Mn) and  four Z-site(Sn) atoms at exactly equal distance (Fig. \ref{cube}(b)). Therefore a difference in the Ni-Mn and Ni-Sn bond lengths obtained from EXAFS analysis conclusively point towards a local structural disorder in the cubic phase.  The Ni-Mn bond length is actually the weighted average distance of Ni from the Y-Mn and the Z-Mn. Ideally, for a L2$_1$ structure, both these contributions should have been identical and equal to the Ni-Sn bond distance. A possible explanation for the discrepancy in the Ni-Mn bond length can be that the two contributions (Ni-[Y-Mn] and Ni-[Z-Mn]) are not equal. That is, there is a slight difference in the bond distance of Y-Mn and the Z-Mn from Ni giving an average Ni-Mn distance shorter than expected. In order to confirm this, Mn EXAFS was fitted again with the $\delta r$'s and $\sigma^2$ varied independently for each path. The resulting fit was excellent as can be seen from Fig. \ref{sn-mnrt} and the final fitted parameters are listed in Table \ref{tab:sn-aus}. It can be seen that the distances obtained from Mn EXAFS anlaysis are indeed lower than the distances calculated from the diffraction data confirming the local disorder in cubic structure. 

\begin{figure}[h]
\epsfig{file=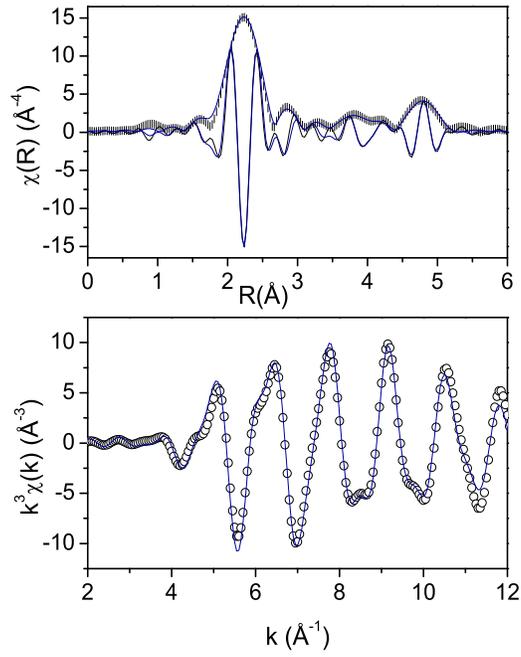, width=8cm, height=10cm}
\caption{\label{sn-mnlt}(Color online) Magnitude and real component of FT of EXAFS spectra in R space and real component of FT in the back transformed k space for Mn K-edge in Ni$_{50}$Mn$_{35}$Sn$_{15}$ obtained at low temperature. Coloured lines represents the fit to the data.}
\end{figure}

\begin{figure}[h]
\epsfig{file=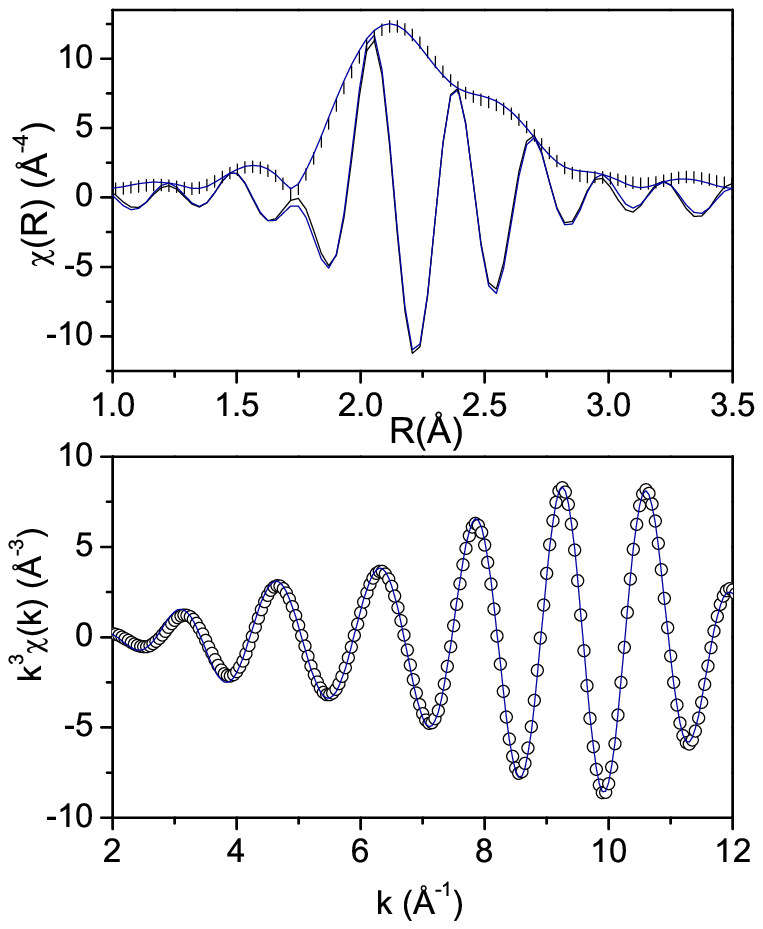, width=8cm, height=10cm}
\caption{\label{sn-nilt}(Color online) Magnitude and real component of FT of EXAFS spectra in R space and real component of FT in the back transformed k space for Ni K-edge in Ni$_{50}$Mn$_{35}$Sn$_{15}$ obtained at low temperature. Fit to the data are shown as coloured lines.}
\end{figure}

The martensitic transformation being in the vicinity of 200 K, EXAFS measured at liquid nitrogen temperature ($\sim$ 77 K) contains no residual cubic component and was used to determine the local structure of the martensitic phase. Neutron diffraction studies by Brown {\it et.al.} \cite{brown18} show that Ni$_{50}$Mn$_{36}$Sn$_{14}$ transforms to a 4{\it O} orthorhombic structure with {\it Pmma} space group in the martesnitic phase. Fig. \ref{cube}(c) represents the {\it Pmma} crystal structure within which a L2$_1$ type sub-cell can be visualized. On comparing this structure with the different scattering paths shown in the L2$_1$ phase of Fig. \ref{cube}(a), it can be clearly understood that with Mn as the central atom, several closely spaced Mn-Ni, Mn-Sn and Mn-Mn bonds are obtained. EXAFS is generally insensitive in resolving such closely spaced scattering paths. Hence, to model the low temperature EXAFS data, these closely spaced scattering paths were grouped into one or two correlations as shown in the Fig. \ref{cube}. The coordination number for the Mn-[Sn/Mn] paths was maintained in accordance with the composition ratio of Mn and Sn. The $ \delta r $s' and $\sigma^2$s' for each of the scattering paths were varied independently without any constraint. Similar fitting procedure was also followed in case of Ni EXAFS. The values given in Table \ref{tab:sn-aus} were found to give a good fit to the experimental spectra as can be seen from Fig. \ref{sn-mnlt} and \ref{sn-nilt}. While an increase of 0.02\AA~ is seen in the Mn-Ni bond distance, the first Mn-Sn and Mn-Mn distances shorten to 2.87\AA~ from those obtained in the austenitic phase. Likewise, the parameters obtained from the analysis of low temperature Ni EXAFS (refer Table \ref{tab:sn-aus}) show that the Ni-Mn distance indeed increases and is in good agreement with the value obtained from Mn EXAFS. The most significant observation here is that the Ni-Sn distance remains unchanged from its room temperature value. This implies that the low temperature modulated structure is due to an unequal movement of constituent atoms with Mn moving with a largest amplitude.

\section{\label{sec:level1}Discussion}
The EXAFS measurements in the cubic and martensitic phase of Ni$_{50}$Mn$_{35}$Sn$_{15}$ bring out an important observation that influence the martensitic transition in this system. Essentially, the present system belongs to the Ni-Mn based ternary Heusler intermetallics having generic formula X$_2$YZ. The stoichiometric Ni$_2$MnSn is a ferromagnet with T$_C$ $\sim$ 360K but does not undergo any martensitic transformation \cite{zaybk}. The exchange interactions in such Heusler alloys is mainly due to the indirect RKKY interaction between Mn atoms mediated by the conduction electrons of the system \cite{klub}. As Mn concentration is increased in the system, martensitic phase transformation sets in and magnetism gets even more complex. The  magnetization results reported in the previous section show that Ni$_{50}$Mn$_{35}$Sn$_{15}$ orders ferromagnetically at T$_C$ = 319 K and undergoes a martensitic transformation $\sim$ 200 K. 

\begin{figure}[h]
\epsfig{file=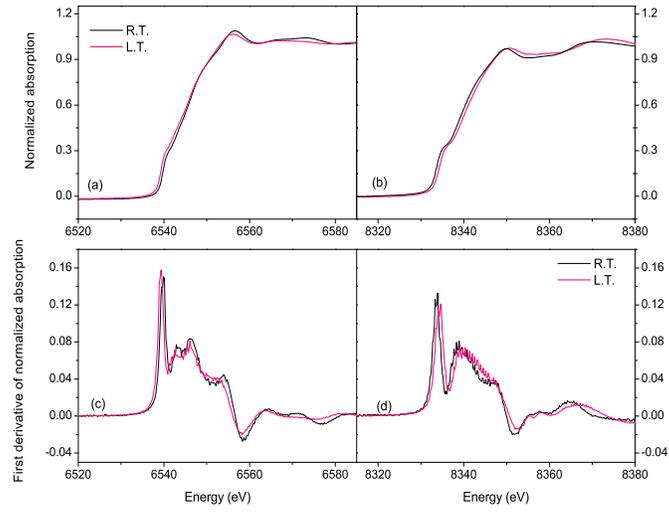, width=10cm, height=8cm}
\caption{\label{xanes}(Color online) Near edge structure at the Mn K-edge (a) and Ni K-edge (b) in the austenitic and martensitic phases in Ni$_{50}$Mn$_{35}$Sn$_{15}$. The first derivative of the near-edge spectra are also shown for Mn K-edge (c) and Ni K-edge (d) at room temperature and low temperature}
\end{figure}

The important result of the EXAFS study is that the local distortions exist in the crystal structure within the cubic framework. These distortions reflect through a shorter Ni-Mn distance and distinctly different Ni-Sn and Ni-Mn bond lengths. The Ni-Mn distance obtained from EXAFS analysis is the average bond length of Ni-[Y-Mn] and Ni-[Z-Mn]. Since the atomic sizes of Sn and Mn are different, a local distortion can occur when Mn atoms occupy the Sn sites. This can lead to a shorter Ni-[Z-Mn] distance as compared to Ni-[Y-Mn] distance. This distortion in the L2$_1$ structure may be one of the factors that influence the martensitic transformation in Ni$_{50}$Mn$_{35}$Sn$_{15}$. A shorter Ni-[Z-Mn] bond implies stronger hybridization of Ni with Z-Mn. The hybridization features between X and Z species of the X$_2$YZ metallic systems is known to influence the binding mechanism \cite{zay-72}. In the case of Ni$_{50}$Mn$_{35}$Sn$_{15}$, the stronger Ni-[Z-Mn] hybridization perhaps results in a redistribution of electrons causing changes in the DOS at Fermi level leading to a martensitic transition. This is further supported by XANES spectroscopy. Figure \ref{xanes} shows the near edge structure at Mn and Ni K edges in austenitic and martensitic phases. It can be clearly seen that, while the Ni-edge shifts by $\sim$0.7 eV to higher energy in the martensitic phase, the Mn edge shifts lower by an exactly similar amount. This indicates that there is a charge transfer from Mn to Ni in the martensitic phase. Such shifts in the edge position have been attributed to hybridization effects dominant near Fermi level for example in hydrogen incorporated Pd films \cite{pdh}. Therefore the exactly opposite shifts in Mn and Ni edge positions imply that in the martensitic phase there is a hybridization between Mn(d) and Ni(d) states near the Fermi level.
 
The martensitic phase change in shape memory alloys is a volume conserving transformation \cite{bhatta}. Therefore the weighted averages of all bond distances obtained from EXAFS analysis in the martensitic phase should be nearly equal to those obtained in the austenitic phase. Indeed this is observed for all distances except for Mn-Sn and Mn-Mn bonds whose lengths shorten from 2.95\AA~ to 2.87\AA.  This indicates that the observed modulated low temperature phase is mainly due to the movement of Mn atoms. The movement of Mn atoms implies that there is a redistribution of Ni-[Y-Mn] and Ni-[Z-Mn] bond distances that give an average Ni-Mn value that is different from that in the cubic phase. Such redistribution results in a scenario wherein the tetrahedral cage of Ni-[Y-Mn] expands giving room for the Ni-[Z-Mn] tetrahedron to shrink and form stronger hybrid states. This situation is analogous to that observed in Ni$_{2+x}$Mn$_{1-x}$Ga alloys \cite{pab}. Here, upon undergoing martensitic transition, the Mn atoms occupying the Y-site move away from Ni while the Ga atoms that occupy the Z-site form stronger bonds with Ni.  

Band structure calculations for X$_2$MnZ Heusler system \cite{klub} have shown that the spin-up {\em 3d} states of Mn are fully occupied and hybridized with the {\em d} states of the X atoms. Whereas the spin down {\em d} states are nearly empty and are almost completely excluded from the Mn site. This results in a localized magnetic moment of Mn. The Z atoms provide electrons to form hybrid states with {\em d} electrons of transition metals and through their valence also determine the degree of occupation of hybrid orbitals. In the case of Ni$_2$MnGa, Zayak {\it et.al} \cite{zay-72} have shown that due to presence of large local magnetic moment of Mn, spin down {\em 3d} electrons of Ni do not find symmetry allowed {\em d} states of Mn to hybridize with. This results in a electrostatic like repulsion between Mn and Ni atoms and leads to a attractive interaction between spin down {\em 3d} electrons of Ni and {\em 4p} electrons of Ga. Such {\em pd} hybridization gives rise to a peak in the density of states that is believed to be responsible for martensitic transformation in this alloy. In Ni$_{50}$Mn$_{35}$Sn$_{15}$, our EXAFS study gives lower average Ni-Mn distances as compared to calculated ones in the austenitic as well as martensitic phase. Further, the XRD profile for Ni$_{50}$Mn$_{35}$Sn$_{15}$ indicates the room temperature austenitic phase to be highly ordered in the L2$_1$ structure. This observation implies that the lower average Ni-Mn distance seen from EXAFS, results from shorter Ni-[Z-Mn] bond. In other words, the Z-Mn is closer to Ni than Y-Mn and Sn atoms. This results in a Ni -{\it d} and Z-Mn -{\it d} hybridization that leads to Jahn-Teller distortion affecting the density of states at E$_F$ and driving the system to a martensitic transition. 

\section{\label{sec:level1}Conclusions}
In conclusion, the analysis of the EXAFS measurements at Ni and Mn K-edges in the austenitic and martensitic phase provide a possible explanation for the occurrence of martensitic transition in Ni$_{50}$Mn$_{35}$Sn$_{15}$. The M(T) data confirms the temperatures of structural and magnetic phase transformations. The results of the EXAFS analysis reveal a local structural disorder in the cubic phase itself that might be the cause for the system to be unstable towards structural transition. Mn atoms are found to move with maximum amplitude from their crystallographic position when forming a modulated martensitic structure. The unequal Ni-Mn bond lengths that results from the presence of excess Mn in the system, shows that it is the {\it dd} hybridization between Ni and Z-Mn that is responsible for martensitic transformation in Ni$_{50}$Mn$_{35}$Sn$_{15}$. 

\acknowledgements
Authors gratefully acknowledge financial assistance from Department of Science and Technology, New Delhi, India and ICTP-Elettra, Trieste, Italy for the proposal 2005743. Thanks are also due to Dr. Luca Olivi for help in EXAFS measurements. P.A.B. would like to thank Council for Scientific and Industrial Research, New Delhi for financial assistance.

\begin{table*}
\caption{Parameters for the cubic and martensitic phases of Ni$_{50}$Mn$_{35}$Sn$_{15}$ determined from the analysis of Mn and Ni K-edge EXAFS recorded at room temperature (RT) and low temperature (LT). The calculated distances for L2$_1$ cell (with $a$ = 5.994 \AA~) are also mentioned. R refers to the bond length and $\sigma^2$ is the thermal mean square variation in the bond length. The analysis were carried out in $k$ range: (2-12)\AA$^{-1}$ with $k$-weight: 3 and $R$ range: (1-5)\AA ~for Mn edge and (1-3)\AA~ for Ni edge. Figures in parentheses indicate uncertainity in the last digit.}
\centering
\begin{ruledtabular}
\begin{tabular}{ccccccc}
\multicolumn{4}{c}{RT - Cubic phase} & \multicolumn{3}{c}{LT - Martensitic phase} \\
Atom and & L2$_1$ Model & & & Atom and & & \\
Coord. No. & R$_{calc}$ (\AA) & R (\AA) & $\sigma^2$ (\AA$^2$) & Coord. No. & R (\AA) & $\sigma^2$ (\AA$^2$)\\
\hline 
\multicolumn{7}{c}{Mn K-edge} \\
Ni1 $\times$ 8 & 2.595 & 2.549(6) & 0.0127(7) & Ni1 $\times$ 8 & 2.568(1) & 0.0077(1) \\
Sn1 $\times$ 3.6 & 2.997 & 2.95(2) & 0.012(2) & Sn1 $\times$ 3.6 & 2.873(3) & 0.0069(3) \\
Mn $\times$ 2.4 & 2.997 & 2.93(1) & 0.008(1) & Mn $\times$ 2.4 & 2.877(4) & 0.0095(5) \\
Mn1 $\times$ 12 & 4.239 & 4.21(2) & 0.021(3) & Mn1 $\times$ 8 & 4.17(2) & 0.023(3) \\
Ni2 $\times$ 24 & 4.970 & 4.95(2) & 0.019(2) & Mn2 $\times$ 4 & 4.291(9) & 0.008(1) \\
MS\footnote{Mn$\rightarrow$Sn3$\rightarrow$Ni1$\rightarrow$Mn} $\times$ 16 & 5.191 & 5.098(3) & 0.014(1) & Ni2 $\times$ 16 & 4.753(6) & 0.0088(7)\\
& & & & Ni3 $\times$ 8 & 4.940(3) & 0.0086(3)\\
\multicolumn{7}{c}{Ni K-edge} \\
Mn1 $\times$ 5.6 & 2.595 & 2.550(7) & 0.0132(9) & Mn1 $\times$ 5.6 & 2.569(2) & 0.0092(2)\\
Sn1 $\times$ 2.4 & 2.595 & 2.601(7) & 0.0072(6) & Sn1 $\times$ 2.4 & 2.607(1) & 0.0035(1)\\
Ni1 $\times$ 6 & 2.997 & 2.98(3) & 0.026(5) & Ni1 $\times$ 2 & 2.83(1) & 0.016(2)\\
& & & & Ni2 $\times$ 4 & 3.15(3) & 0.027(5) \\
\end{tabular}
\end{ruledtabular}
\label{tab:sn-aus}
\end{table*}

\end{document}